# Discovering Attractive Products based on Influence Sets

Anastasios Arvanitis [#1], Antonios Deligiannakis [*2]

[#]*National Technical University of Athens, Greece*    [*]*Technical University of Crete, Greece*
[1]`anarv@dblab.ntua.gr`  [2]`adeli@softnet.tuc.gr`

*Abstract*— Skyline queries have been widely used as a practical tool for multi-criteria decision analysis and for applications involving preference queries. For example, in a typical online retail application, skyline queries can help customers select the most interesting, among a pool of available, products. Recently, reverse skyline queries have been proposed, highlighting the manufacturer's perspective, i.e. how to determine the expected buyers of a given product. In this work we develop novel algorithms for two important classes of queries involving customer preferences. We first propose a novel algorithm, termed as RSA, for answering reverse skyline queries. We then introduce a new type of queries, namely the $k$-Most Attractive Candidates (k-MAC) query. In this type of queries, given a set of existing product specifications $P$, a set of customer preferences $C$ and a set of new candidate products $Q$, the k-MAC query returns the set of $k$ candidate products from $Q$ that jointly maximizes the total number of expected buyers, measured as the cardinality of the union of individual reverse skyline sets (i.e., *influence sets*). Applying existing approaches to solve this problem would require calculating the reverse skyline set for each candidate, which is prohibitively expensive for large data sets. We, thus, propose a batched algorithm for this problem and compare its performance against a branch-and-bound variant that we devise. Both of these algorithms use in their core variants of our RSA algorithm. Our experimental study using both synthetic and real data sets demonstrates that our proposed algorithms outperform existing, or naive solutions to our studied classes of queries.

## I. INTRODUCTION

Skyline queries have been widely recognized as a practical tool in many real world applications, such as multi-criteria decision analysis and applications involving preference queries. Prior work mainly focuses on helping customers select the 'best', based on their preferences, product among a pool of available products. Recently, reverse skyline queries have been proposed [6], highlighting the manufacturer's perspective, i.e., how to determine the expected buyers of a given product. The state of the art algorithm for answering reverse skylines queries is the BRS algorithm, proposed in [24].

In this work we develop novel algorithms for two important classes of queries involving customer preferences. We first propose a novel algorithm, termed as RSA, for reverse skyline queries. Our RSA algorithm significantly reduces both the I/O cost, as well as the computation cost, when compared to the state-of-the-art reverse skyline algorithm BRS, while at the same time being able to quickly report the first query results.

We then introduce a new type of queries, namely the $k$-Most Attractive Candidates (k-MAC) queries. In this type of queries, given a set of existing product specifications $P$, a set of customer preferences $C$ and a set of new candidate products $Q$, the k-MAC query returns the set of $k$ candidate products from $Q$ that jointly maximizes the total number of expected buyers, measured as the cardinality of the union of individual reverse skyline sets (i.e., *influence sets*). Two examples of real world applications that may benefit from answering reverse skyline and k-MAC queries, respectively, are the following:

**Market Analysis.** The marketing department of a laptop manufacturer wants to discover which are the most "attractive" laptops in the market, i.e. which laptops satisfy the preferences of the largest number of customers. Let us assume that the manufacturer has assembled a set $P$ of all laptops available in the market. Let us further assume that the manufacturer has conducted a market survey and compiled a list of customers' preferences $C$ over a laptop's attributes such as screen size, CPU, operating system etc. To enable more effective promotion of its own products to interested customers and adapt their advertising strategy, the marketing department needs to discover the potential buyers for the company products.

**Product Placement.** Consider a mobile carrier operator that wants to introduce a set of new cell phone plans. The operator maintains a database of existing plans, customer statistics (i.e., voice usage duration, number of text messages, data volume consumed per month, etc.) and a list of new phone plans under consideration. The company is looking for the set of phone plans that jointly would have the largest possible number of potential subscribers.

Prior work [19] has tackled a similar to k-MAC problem in the context or top-k queries. However, this approach requires learning and precisely formulating user preference functions as weight vectors, which is rather artificial and impractical in real life. On the other hand, in our work each user is allowed to formulate his preferences directly in terms of the desired product specifications. Moreover, in contrast to our k-MAC query, the total profitability of the query result in [19] is not guaranteed because multiple products might be attractive for the same potential buyers.

Applying single point skyline approaches to solve the k-MAC problem would require calculating the reverse skyline set for each candidate, which is prohibitively expensive for large data sets. We, thus, propose a batched algorithms for the k-MAC problem and compare its performance against a branch-and-bound variant that we devise. Both of these algorithms use in their core variants of our RSA algorithm. In brief, the contributions of this paper are:

- We identify potential drawbacks, which are more evident in

high-dimensional data where the number of skyline entries is typically larger, of the BRS algorithm. Based on our observations, we then introduce a novel algorithm RSA for (single) reverse skyline query evaluation.

- We introduce a new type of skyline query, termed as the $k$-Most Attractive Candidates (k-MAC) query; k-MAC selects the set of $k$ candidate products that jointly maximizes the total number of expected buyers.
- We propose a batch algorithm for answering k-MAC queries and compare against a branch-and-bound alternative that we devise. Both of these algorithms utilize in their core our RSA algorithm, but significantly improve upon naive techniques that process all candidate products individually.
- We perform an extensive experimental study using both synthetic and real data sets. Our study demonstrates that (*i*) our RSA algorithm outperforms the BRS algorithm, for the reverse skyline problem, in terms of I/O, CPU cost and progressiveness of the output, and (*ii*) that our proposed batch algorithm outperforms naive approaches that process each candidate individually.

This paper proceeds as follows. Section II reviews background information and definitions, while also describing the current state-of-the-art BRS algorithm for computing reverse skylines. Section III presents our RSA algorithm and Section IV introduces the k-MAC queries. In Section VI query evaluation, and in Section V we present a greedy method for discovering the best $k$ candidate products. Section VII provides our experimental results. Section VIII reviews related work, while Section IX presents concluding remarks.

## II. PRELIMINARIES

In this section we first provide some necessary definitions in Section II-A. In Section II-B we then present the notion of influence sets and properties regarding influence sets, while in Section II-C we describe the state of the art BRS algorithm for computing influence sets. Table I summarizes the most frequently used notation of this paper. The corresponding definitions are presented in appropriate areas of the text.

### A. Single Point Reverse Skylines

Consider two sets of points, denoted as $P$ and $C$, in the same $D$-dimensional space. We use lower-case symbols to denote an item that belongs to the corresponding upper-case set, e.g., $p \in P$. We refer to each point $p \in P$ as a *product*. Each product is a multi-dimensional point, with $p_i$ denoting the product's value for dimension (attribute) $A_i$. For example, assuming that products are notebooks, the dimensions of $p_i$ may correspond to the notebook's price, weight, CPU speed, screen size, etc. Further, each point $c \in C$ represents a customer's preferred notebook specifications that he would be interested in; we refer to each point $c$ as a *customer*.

Clearly, for some of the dimensions there exists a clear preference (i.e., an optimal value) for all customers. In our notebook scenario, price and weight should be as low as possible (amongst two notebooks with all the other characteristics being identical, it is conceivable that customers would prefer the most economic/light alternative), whereas RAM size should be as large as possible. However, note that there is no objectively optimal value for CPU speed and screen size. A large screen is more practical but it also compromises portability. Similarly, faster CPUs increase heat and fan noise and may decrease battery life. Thus, while one customer might prefer a notebook with 13" screen, another customer might prefer one with a 15,4" screen. In general, customers are more interested to the products that are closer to their preferences. In order to capture the preferences of a customer $c$ the notion of dynamic dominance among products has been introduced.

TABLE I
NOTATION

| Symbol | Definition |
|---|---|
| $P, C, Q$ | set of products, customers, candidate products |
| $p_i \prec_c p_j$ | product $p_i$ dominates $p_j$ with respect to customer $c$ |
| $m_i(q)$ | midpoint of $p_i$ with respect to $q$ |
| $T_P, T_C, T_C^A$ | R-tree of products, R-tree of customers, aR-tree of customers |
| $e_p, e_c$ | an entry corresponding to a node of $T_P, T_C$ |
| $e_x^-(q), e_x^+(q)$ | min-corner and max-corner of $e_x$ (w.r.t. $q$) |
| $SKY(c)$ | skyline with respect to customer $c$ |
| $RSKY(q)$ | influence set of candidate $q$ |
| $IS(q)$ | influence score of candidate $q$ |
| $IR(q)$ | influence region of candidate $q$ |
| $E_P(q), E_C(q)$ | priority queue on active entries for $q$ |
| $L, U$ | set of min-corners, minmax-corners for $E_P(q)$ |
| $IS^-(q), IS^+(q)$ | lower and upper bound of $IS(q)$ |
| $IR^-(q), IR^+(q)$ | lower and upper bound of $IR(q)$ |
| $RSKY(Q)$ | joint influence set of $Q$ |
| $IS(Q)$ | joint influence score of $Q$ |

*Definition 1:* (Dynamic Dominance) (from [6]): Let $c \in C$, $p, p' \in P$. A product $p$ dynamically dominates $p'$ with respect to $c$, denoted as $p \prec_c p'$, iff for each dimension $|p_i - c_i| \leq |p'_i - c_i|$ and there exists at least one dimension such that $|p_i - c_i| < |p'_i - c_i|$.

Note that this definition can accommodate dimensions with objectively optimal values (static dominance) where smaller (larger) values are preferred by simply setting $c_i$ to the minimum (resp. maximum) value of dimension $A_i$. For example, assuming that lighter notebooks are preferred, we can simply set for all customers $c_{weight} = 0$.

*Definition 2:* (Dynamic Skyline) (from [6]): The dynamic skyline with respect to a customer $c \in C$, denoted as $SKY(c)$, contains all products $p \in P$ that are not dynamically dominated with respect to $c$ by any other $p' \in P$.

Consider a set of existing products $P = \{p_1, p_2, p_3, p_4\}$ and customers $C = \{c_1, c_2, c_3\}$. Figure 1(a) illustrates the dynamic skyline of $c_1$ that includes notebooks $p_2$ and $p_4$, in a sample scenario of 2-dimensional preferences that reference the CPU speed and the screen size of a notebook. Points in the shaded areas are dynamically dominated by points belonging to the dynamic skyline of $c_1$. Notice that, since we are interested in the absolute distance between products, a product might dominate other products that belong to different quadrants with respect to a customer. For example, $p_1$ and $p_3$ in the upper right quadrant are dynamically dominated by $p_2$ in the lower right quadrant because $p_2$ has a CPU speed and a screen size that are both closer to $c_1$ than the corresponding characteristics of $p_1$ and of $p_3$. Figures 1(b) and 1(c) illustrate the dynamic

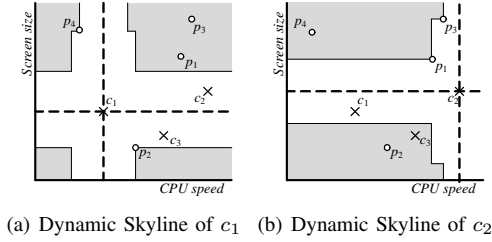
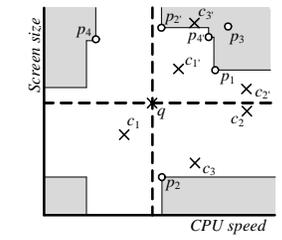
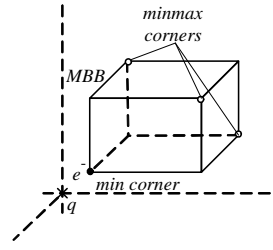
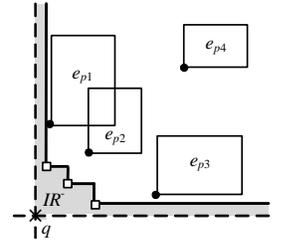

(a) Dynamic Skyline of $c_1$  (b) Dynamic Skyline of $c_2$

(a) Transformed space to $\Omega_0$ with respect to $q$

(a) Example MBB

(b) Lower bound for IR($q$)

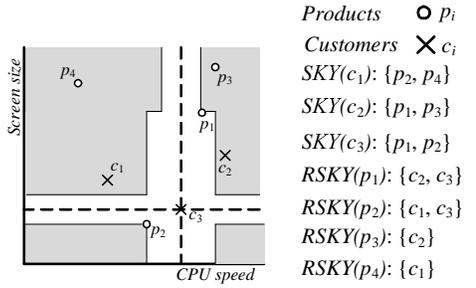
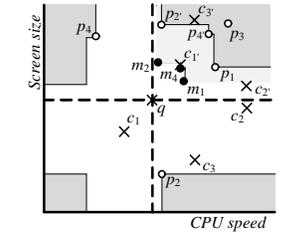
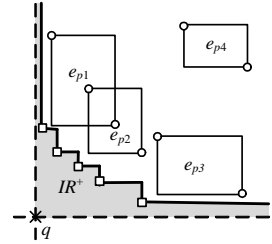
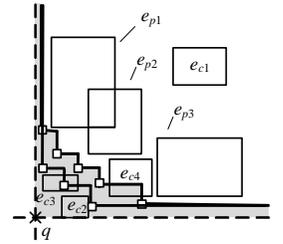

Products ○ $p_i$
Customers × $c_i$
$SKY(c_1)$: $\{p_2, p_4\}$
$SKY(c_2)$: $\{p_1, p_3\}$
$SKY(c_3)$: $\{p_1, p_2\}$
$RSKY(p_1)$: $\{c_2, c_3\}$
$RSKY(p_2)$: $\{c_1, c_3\}$
$RSKY(p_3)$: $\{c_2\}$
$RSKY(p_4)$: $\{c_1\}$

(c) Dynamic Skyline of $c_3$  (d) Dynamic Skylines and Influence Sets

Fig. 1. Dynamic Skylines example

(b) Midpoints with respect to $q$

Fig. 2. Influence region of $q$

(c) Upper bound for IR($q$)

(d) $e_c$ w.r.t $E_P$

Fig. 3. Influence Regions

skylines of customers $c_2$ and $c_3$ respectively.

Now we highlight the product's perspective by introducing the definition of *bichromatic reverse skylines*.

*Definition 3:* (Bichromatic Reverse Skyline) (from [15]): Let $P$ be a set of products and $C$ be a set of customers. The bichromatic reverse skyline of $p$, denoted as $RSKY(p)$ contains all customers $c \in C$ such that $p \in SKY(c)$.

In other words, the bichromatic reverse skyline of a product $p$ contains all customers $c$ that find $p$ as 'attractive'. In the following we will refer to the bichromatic reverse skyline of $p$ as the *influence set* of $p$. Figure 1(d) illustrates the influence sets of products $p_1$, $p_2$, $p_3$ and $p_4$.

Clearly, the cardinality of the influence set $RSKY(p)$ is a useful metric of the product's impact in the market. We will refer to $|RSKY(p)|$ as the *influence score* IS($p$). In our example IS($p_1$) = 2, IS($p_2$) = 2, IS($p_3$) = 1 and IS($p_4$) = 1.

### B. Influence Region

Consider a new product $q$. The new product partitions the $D$-dimensional space into $2^D$ orthants $\Omega_i$, each identified by a number in the range $[0, 2^D - 1]$. For example, in Figure 2 where $D = 2$, $q_1$ partitions the space in 4 orthants (quadrants). Since all orthants are symmetric and we are interested in the absolute distance between products, we can map all products to $\Omega_0$ as illustrated in Figure 2(a). For simplicity, we hereafter concentrate on $\Omega_0$ with respect to a query point $q$.

For every dynamic skyline point $p_i$, let $m_i(q)$ be the midpoint of the segment connecting a query point $q$ with $p_i$. In Figure 2(b) black points $m_1$, $m_2$ and $m_4$ represent the midpoints of $p_1$, $p_2$ and $p_4$ with respect to $q$.

In the rest of this work, in order to get rid of the complication of maintaining both points and midpoint skylines, whenever we refer to a product $p_i$ we will imply the corresponding $m_i(q)$ in respect to $q$. Further, we will assume that each dynamic skyline point $p_i$ with respect to $q$ is mapped to its midpoint skyline $m_i$ on the fly.

The *influence region* of a query point $q$, denoted as IR($q$), is the union of all areas not dynamically dominated (*anti-dominance areas*) with respect to $q$ by the midpoint skylines of $q$. The area in $\Omega_0$ that is not shaded in Figure 2(b) draws the influence region for $q$.

*Lemma 1:* (from [15]) A customer $c$ belongs to the influence set $RSKY(q)$ of $q$ iff $c$ lies inside the influence region of $q$ i.e., $c \in$ IR($q$) $\Leftrightarrow c \in RSKY(q)$.

Returning to the example of Figure 2(b), notice that only $c_2$ lies inside IR($q$). Therefore, $RSKY(q) = \{c_2\}$.

In our discussion hereafter, we assume that all points (either products or customer preferences) are indexed using a multi-dimensional index (e.g. R-trees, kd-trees etc.). Although any multidimensional index can be used instead, R-trees represent an attractive choice because of their popularity in commercial database systems. Figure 3(a) shows an example minimum bounding box (MBB) $e$.

Inside each MBB $e$, let *min-corner* $e^-(c)$ denote that point (in $e$) at the minimum distance from a query point $q$[1]. The min-corner dominates the largest possible part of the orthant. Further, due to the construction of an MBB, i.e., it is the smallest possible box that encloses a set of points, an MBB has the property that each face (i.e., side) must contain at least one point (otherwise the box would be shrunk). The points that reside in each of the $d$ faces closest to $q$ and are the farthest from the origin $q$ are denoted as *minmax-corners*. Each MBB contains $D$ minmax-corners. Independently of how products within $e$ are distributed, any point in $e$ certainly dominates the area that the minmax-corners do, while at best it dominates the area that the min-corner does.

Given a set of MBBs, we can derive two sets: the set of all min-corners denoted as $L$ and the set of all minmax-corners denoted as $U$ with respect to $q$. Figure 3(b) presents an example, assuming $E_P = \{e_{p_1}, e_{p_2}, e_{p_3}, e_{p_4}\}$, where $e_{p_i}$ denotes a product entry. In Figures 3(b), 3(c) black and hollow circles represent the min-corners and minmax-corners respectively and rectangles represent midpoints.

---

[1]Note that the notation does not use $q$

Consider the example of Figure 3(b). The grey area represents a lower bound of the actual influence region $IR^-$ and it is defined as the space not dominated by any min-corner $l \in L$. Continuing the example, in Figure 3(c) the grey area represents an upper bound of the actual influence region $IR^+$, defined as the space not dominated by any minmax-corner $u \in U$. It follows [24]:

*Lemma 2:* If an entry $e_c$ is dominated by any $u \in U$, i.e. $e_c$ is completely outside $IR^+(q)$, $e_c$ certainly cannot contain any customer belonging to $IR(q)$. Consequently, according to Lemma 1, $e_c$ does not contribute any reverse skyline point, hence it can be pruned.

For example, $e_{c1}$ in Figure 3(d) can be pruned because it is completely outside $IR^+(q)$.

### C. The BRS Algorithm

In the following we detail the state-of-the-art *Bichromatic Reverse Skyline* (BRS) algorithm [24] that efficiently calculates the influence set of a single query point $q$. BRS aims at minimizing the I/O cost (*i*) by progressively refining the influence region of $q$ until the influence set of $q$ has been retrieved, (*ii*) by applying Lemma 2 to prune $e_c$ entries that do not contribute to $RSKY(q)$.

BRS uses two indexes, an R-tree $T_P$ on the set of products $P$ and another $T_C$ on the set of customers $C$. Initially, the algorithm inserts all root entries of $T_P$ (resp. $T_C$) in a priority queue $E_P$ (resp. $E_C$) sorted with the minimum Euclidean distance (*mindist*) of each entry from $q$. BRS extracts a set $L$ of all min-corners and a set $U$ of all minmax-corners of $e_p \in E_P$. Further, in order to reduce the number of subsequent dominance checks, BRS calculates the skylines of $L$ and $U$, denoted as $SKY(L)$ and $SKY(U)$ respectively.

In each iteration BRS expands the entry in $E_P$ with the minimum mindist from $q$ and updates the current $L$ and $U$ and their skylines $SKY(L)$ and $SKY(U)$. Then, all $e_c \in E_C$ are checked for dominance with $SKY(L)$ and $SKY(U)$. If $e_c$ is not dominated by $SKY(L)$ (i.e. it intersects $IR^-(q)$), BRS expands $e_c$ as it will certainly contain at least one customer belonging to $IR(q)$. Returning to Figure 3(d), $e_{c3}$ intersects $IR^-(q)$, meaning that it contains at least one customer inside $IR(q)$, therefore it is expanded. In contrast, if $e_c$ (such as $e_{c1}$ in Figure 3(d)) is dominated by $SKY(U)$, then $e_c$ can be safely pruned according to Lemma 2.

Additionally, in each iteration BRS checks all $e_p \in E_P$ with the current $SKY(U)$. If an entry $e_p$ is dominated by any $u \in SKY(U)$, then $e_p$ can be removed from $E_P$, as it cannot contribute to further refining $IR(q)$. For example, $e_{p4}$ in Figure 3(c) can be pruned because it is dominated by at least one minmax-corner $u \in SKY(U)$. BRS terminates when $E_C$ becomes empty, i.e. the position of all customers either inside or outside $IR(q)$ has been determined.

## III. THE RSA ALGORITHM

In this section, we detail the drawbacks of BRS and then propose our more efficient reverse skyline algorithm, RSA.

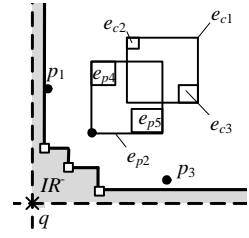

Fig. 4. Expanding $e_{p_2}$ entry in BRS is redundant, as it cannot refine $e_{c_1}$

### A. BRS Shortcomings

**Complexity Analysis.** Let $p_k$, $c_k$ denote the sizes of active lists $E_P$ and $E_C$, respectively, after the $k$-th iteration of the BRS algorithm. The worst-case cardinality of $p_k$ and $c_k$ are $|P|$ and $|C|$ respectively. In each iteration, the BRS algorithm maintains both $SKY(L)$ and $SKY(U)$, two sets with $O(|P|)$ and $O(D|P|)$ entries respectively, where $D$ is the dimensionality of the data set. BRS then checks for dominance each entry in $E_P$ and $E_C$ with $SKY(L)$ and $SKY(U)$. Thus, each iteration entails $O(D^2|P| \times (|P| + |C|))$ comparisons[2].

Clearly, the processing cost of BRS depends heavily on the size of the intermediate $SKY(L)$ and $SKY(U)$ sets. Godfrey [8] shows that for a uniformly distributed data set, the size of the skyline set is $\Theta(\frac{(lnN)^{D-1}}{(D-1)!})$, where $N$ is the data set cardinality. Thus, for higher dimensional data, or in larger data sets, the processing cost of maintaining the lower and upper bounded skylines becomes prohibitively expensive. As we also demonstrate in our experiments, BRS becomes impractical when $N \geq 10^6$ or $D \geq 4$. Motivated by the above analysis, we seek to develop in this section a more efficient and scalable reverse skyline algorithm, which will eliminate the dependency on the $SKY(L)$ and $SKY(U)$ sets, thus being able to handle high dimensional data, or, in general, data where the size of (reverse) skylines is large.

**Visiting Order.** BRS algorithm performs a synchronous traversal on the $T_P$ and $T_C$ indexes, which are built on product and customer points, respectively. The execution of the BRS algorithm follows a monotonic order based on the Euclidean distance of $e_p$ entries from $q$. This visiting order ensures that the number of I/O operations on $T_P$ is minimized. However, it cannot guarantee that the total number of I/Os (both on $T_P$ and $T_C$) is optimal.

We argue that, depending on the relative position of $e_p$ and $e_c$ with respect to $q$, a different visiting order might entail fewer I/Os than those required by the BRS algorithm. Fig. 4 illustrates one such scenario, where the nodes $e_{p_2}$ and $e_{c_1}$ have not been expanded yet. In this scenario, BRS would proceed by expanding $e_{p_2}$, thus revealing the entries $e_{p_4}$ and $e_{p_5}$. Unfortunately, $e_{c_1}$ is not affected by this refinement and it still has to be accessed. However, if we first expand $e_{c_1}$, this operation would reveal $e_{c_2}$ and $e_{c_3}$, which can be pruned by $p_1$ and $p_3$ respectively, eliminating the need to access $e_{p_2}$. Clearly, in this scenario the I/O operation on $e_{p_2}$ was redundant. In this work, we introduce a new reverse skyline algorithm that follows a visiting order based on a customer entry's tree level (primary sort criterion) and Euclidean distance (secondary sort criterion) from $q$. Following the proposed visiting order, a product entry will be accessed only as long as it is necessary in order to determine if a customer point belongs to $RSKY(q)$, which leads to fewer total I/O operations on both indexes.

---
[2]Recall that each dominance check requires $D$ comparisons

**Progressiveness.** BRS algorithm iteratively refines $IR^-(q)$ and reports the customer points that lie inside $IR(q)$. In order to retrieve the first result, several iterations of BRS might be required, which is a not desirable property for applications that require a quick response that contains only a fraction of the total output. We thus seek to develop an algorithm that reports the first results faster than BRS algorithm and which is more progressive in terms of the fraction of output reported.

*B. RSA Algorithm*

We now propose our RSA (Reverse Skyline Algorithm), which aims to eliminate the shortcomings outlined above. In particular, our RSA algorithm:

- Does not require the maintenance of the $SKY(L)$ and $SKY(U)$ sets, and is, therefore, significantly less expensive in terms of processing cost.
- It checks one customer entry per iteration following a visiting order based on the entry's tree level (primary sort criterion) and Euclidean distance (secondary sort criterion) from $q$.
- It accesses a $e_p$ entry only if it is necessary in order to determine if a customer point belongs to $RSKY(q)$.

**Data Structures Used and Basic Intuition.** RSA maintains five data structures for its operation:

- A priority queue $E_P$ on the set of active products entries
- A priority queue $E_C$ on the set of active customer entries
- A R-tree $T_P$ on the set of products $P$
- A R-tree $T_C$ on the set of customers $C$
- A set $SKY(q)$ of the currently found midpoint skylines

The two priority queues are sorted based on a dual sorting criterion: primarily, based on the tree level of the stored entries and, subsequently, using the mindist of each entry from $q$. Thus, leaf entries are given higher priority and are processed first, while the examination of non-leaf entries is postponed as much as possible. By first processing all leaf $e_c$ entries, the algorithm may reveal a midpoint skyline, which will be subsequently used to prune a non-leaf $e_c$, thus avoiding an access on $T_C$. The same intuition holds for $e_p$ entries as well; when a $e_c$ entry is checked for dominance with $E_P$, first all leaf $e_p$ entries will be examined. As long as no non-leaf $e_p$ dominates $e_c$, only then will RSA proceed to expand the nearest non-leaf $e_p$ entry to $q$. This change in the visiting order of $E_P$ reduces the number of accesses on $T_P$ as well.

**Algorithm Description.** Now, we describe the execution of our RSA algorithm, while providing its detailed pseudocode in Alg. 1. Initially, RSA inserts all root entries of $T_P$ (resp. $T_C$) in the priority queue $E_P$ (resp. $E_C$). Further, RSA maintains a set $SKY(q)$ of the currently found midpoint skylines which are used for pruning based on Lemma 1. RSA proceeds in iterations. In each iteration RSA extracts the entry in $E_C$ having the minimum key from $q$ (Line 5) and checks the following pruning conditions:

**Algorithm 1: RSA**

**Input**: $q$ a query point, $T_P$ R-tree on products, $T_C$ R-tree on customers, $E_P(q)$ priority queue on product entries, $E_C(q)$ priority queue on customer entries
**Output**: $RSKY(q)$ reverse skylines of $q$
**Variables**: $SKY(q)$ currently found midpoint skylines of products w.r.t $q$

1 **begin**
2     $SKY(q) := \varnothing; RSKY(q) := \varnothing;$
3     **while** $E_C \neq \varnothing$ **do**
4         dominated := false;
5         $E_C(q).pop() \to e_c;$
6         **if** *dominated*($e_c$, $SKY(q)$) **then**
7             dominated := true;
8             continue;
9         **if** $e_c$ *is a non-leaf entry* **then**
10            EntryExpand($e_c$, $SKY(q)$, $E_C(q)$, $T_C$);
11         **else**
12            **foreach** $e_p \in E_P(q)$ **do**
13                midpoint($e_p, q$) $\to m$;
14                **if** $e_c$ *is dominated by $m$* **then**
15                   **if** $e_p$ *is a leaf entry* **then**
16                       **if** *(dominated(m, $SKY(q)$) == false)* **then**
17                           $SKY(q).push(m)$;
18                   dominated := true;
19                   break;
20                **else**
21                   EntryExpand($e_p$, $SKY(q)$, $E_p(q)$, $T_P$);
22                $E_P(q).remove(e_p)$;
23         **if** *(dominated == false)* **then**
24            $RSKY(q).push(e_c)$;
25     **return** $RSKY(q)$;

1. If $e_c$ is dominated by any point that belongs in the currently found midpoint skylines $SKY(q)$, $e_c$ can be removed from $E_C$ based on Lemma 1 (Lines 6-8).

2. Otherwise, if $e_c$ is a non-leaf entry (Line 9), $e_c$ is expanded and children nodes are inserted into $E_C$ (Line 10 and Alg. 2, described later in this section).

3. Else, for all $e_p$ entries in $E_P$ (Lines 12-22):

- If $e_c$ is dominated by the midpoint of a leaf entry $e_p \in E_P$ (Line 15), then $e_c$ can be removed from $E_C$, based on Lemma 1, and the midpoint of $e_p$ is inserted into $SKY(q)$ (Line 17)).

- Else if $e_c$ is dominated by the midpoint of the *min-corner* $e_p^-$ of a non-leaf $e_p \in E_P$, $e_p$ is expanded and its children entries are inserted into $E_P$ (Line 21).

Finally, if $e_c$ has not been pruned by any of the above conditions, then $e_c$ is a reverse skyline point (Lines 23-24) and can be at that stage directly reported in the result. The RSA algorithm terminates when $E_C$ becomes empty and then $RSKY(q)$ is returned (Line 25).

**Optimizations.** Observe that any product entry that is already dominated w.r.t. to $q$ by any currently found skyline point, will definitely not contribute in further refining the $RSKY(q)$. Consequently, it can be safely pruned from further processing. We apply this observation as an additional optimization step whenever expanding a product entry. The detailed pseudocode is provided in Alg. 2.

In particular, whenever a $e_p$ ($e_c$) entry is expanded, before inserting children nodes to $E_P$ ($E_C$), the RSA algorithm performs the following: First a temporary queue is created and children nodes are iteratively pushed inside the queue

**Algorithm 2:** EntryExpand

**Input**: $e_x$ a product or customer entry, $SKY(q)$ currently found midpoint skyline set w.r.t to $q$, $E_x(q)$ priority queue on $e_x$, $T_x$ index on $e_x$
**Output**: $E_x(q)$
**Variables**: $L$ queue on children of $e_x$

1 **begin**
2    $L := \varnothing$;
3    **foreach** $e_{x_l}$ children of $e_x$ **do**
4       **if** $e_x$ is a product entry **then**
5          **if** (dominated($e_{x_l}$, $L$) == true) **then**
6             continue;
7       **if** (dominated($e_{x_l}$, $SKY(q)$) == false) **then**
8          $L.push(e_{x_l})$;
9    **foreach** $e_{x_l} \in L$ **do**
10       $E_x(q).push(e_{x_l})$;
11    **return** $E_x(q)$;

(Line 8). When expanding a customer entry, each child node is checked only with the currently found $SKY(q)$ (Line 7). If it is dominated, then the child node can be pruned according to Lemma 1. On the other hand, when expanding a product entry $e_p$, before pushing $e_p$'s children nodes inside the temporary queue, each children node is checked for dominance w.r.t to $q$ with: (i) children nodes already inside the queue (Line 5), and (ii) the currently found $SKY(q)$ (Line 7). Finally, after all qualified children nodes have been determined they are pushed inside $E_P$ ($E_C$) (Line 10).

This optimization results in fewer children nodes pushed inside the priority queues $E_P$ ($E_C$) which leads to: (i) less space needed for $E_P$ ($E_C$) in intermediate iterations, and (ii) reduced processing time for subsequent dominance checks, since reducing the number of pushed entries results in fewer entries being processed in the future and in smaller data structures being used.

## IV. PROBLEM FORMULATION: k-MAC QUERIES

After presenting our RSA algorithm for efficiently calculating the reverse skyline set for a single candidate, in this section we introduce the notion of $k$-Most Attractive Candidates (k-MAC) queries. We first present a motivating sample scenario, which highlights the importance of k-MAC queries. We then formally define our k-MAC problem.

**Motivating Scenario.** A laptop manufacturer wants to produce $k$ new notebooks, among a set of feasible candidate laptop configurations $Q$ proposed by the engineering department. In this process, the manufacturer needs to consider three sets: (1) the existing competitor products $P$, (2) the set of customers' preferred specifications $C$, and (3) a set of candidate products $Q$. We will refer to each $q \in Q$ as a *candidate*.

Clearly, the goal of the manufacturer is to identify the specifications that are expected to *jointly* attract the largest number of potential buyers. Please note that this is different than simply selecting the $k$ products that are the most attractive individually, since it does not make much sense to select products that seem attractive to the same subset of customers.

**Problem Definition.** We first define the *joint influence set* for a set of candidates $Q$. We then define the notion of the *joint influence score* and introduce the *k-Most Attractive Candidates (k-MAC) query*.

*Definition 4:* (Joint Influence Set): Given a set of products $P$, a set of customers $C$ and a set of candidates $Q$, the joint influence set of $Q$, denoted as $RSKY(Q)$, is defined as the union of individual influence sets of any $q_i \in Q$.

$$RSKY(Q) = \bigcup_{q_i \in Q} RSKY(q_i)$$

Following the above definition, the *joint influence score* $IS(Q)$ for a set of candidates $Q$ is equal to the size ($|RSKY(Q)|$) of the joint influence set of $Q$.

We now introduce the *k-Most Attractive Candidates* (k-MAC) query as follows:

*Definition 5:* $k$-Most Attractive Candidates (k-MAC) query): Given a set of products $P$, a set of customers $C$, a set of candidates $Q$ and a positive integer $k > 1$, determine the subset $Q_i \subseteq Q$, such that $|Q_i| = k$ and the joint influence score $IS(Q_i)$ of $Q_i$ is maximized.[3]

Obviously, the 1-MAC problem returns the most attractive candidate. Returning to Figure 1, assuming that the four products represent the four candidates, the solution to a 2-MAC query could return as solution either the set $\{p_1, p_2\}$, or the set $\{p_1, p_4\}$ or $\{p_2, p_3\}$, all of them having a joint influence score of 3. In case of ties, these ties can be resolved arbitrarily.

Unfortunately, answering k-MAC queries is non-trivial. This problem can trivially be reduced to the more general *maximum k-coverage* problem. Thus, even if we consider the much simpler problem where all the influence sets of all candidates have been computed, an exhaustive search over all potential $k$ cardinality subsets of $Q$ is NP-hard.

## V. A GREEDY COMPONENT

Based on the complexity of computing the subset of $k$ products, we now seek an efficient, greedy algorithm for this problem. Our solution is based on the generic $k$-stage Greedy Selection Algorithm (kGSA) provided in [10], developed for finding efficient approximate solutions to the maximum k-coverage problem.

*Lemma 3:* (from [10]) kGSA returns an approximate solution to the maximum $k$-coverage problem that is guaranteed to be within a factor $1 - 1/e$ from the optimal solution.

We now show how we can adapt kGSA for the k-MAC problem. The corresponding pseudocode appears in Alg. 3. kGSA takes as input a set of candidate products $Q$ and their associated influence sets and returns a set $Q' \subseteq Q$, $|Q'| = k$ that contains the candidates which formulate a $1 - \frac{1}{e}$ approximate solution to the k-MAC query. kGSA proceeds in iterations, by adding one candidate into $Q'$ during each iteration. All candidates are examined (Lines 5-8) at each iteration, and kGSA selects the one that, if added in $Q'$, results in the largest increase of the influence score of $Q'$. This best candidate is maintained in the variable $best\_cand$ (Line 8), and its resulting reverse skyline set is maintained in the variable $tempRSKY$ (Line 7). kGSA terminates after $k$ iterations and returns $Q'$.

---
[3]Note that since the joint influence score of a set of candidates is non-decreasing when adding new members in the set, the optimal solution in this problem is guaranteed to contain exactly $k$ candidates.

**Algorithm 3:** $k$-stage Greedy Selection Algorithm

**Input**: $Q$ a set of candidates, $RSKY(q_i)$ reverse skylines of $q_i$, $k$
**Output**: $Q'$ the most attractive set of candidates where $|Q'| = k$

```
1  begin
2      Q' := ∅; tempRSKY := ∅; maxRSKY := ∅;
3      while |Q'| < k do
4          tempRSKY := maxRSKY;
5          foreach q_i ∈ Q do
6              if |RSKY(q_i) ⋃ maxRSKY| > |tempRSKY| then
7                  tempRSKY := RSKY(q_i) ⋃ maxRSKY;
8                  best_cand := {q_i};
9          maxRSKY := tempRSKY;
10         Q := Q − best_cand;
11         Q' := Q' ⋃ best_cand;
12     return Q';
```

**Algorithm 4:** Basic Algorithm

**Input**: $Q$ a set of candidates, $T_P$ R-tree on products, $T_C$ R-tree on customers
**Variables**: $E_P(q_i)$ set of active product entries for $q_i$, $E_C(q_i)$ set of active customer entries for $q_i$, $RSKY(q_i)$ reverse skylines for $q_i$, $SKY(q_i)$ midpoint skylines of $q_i$

```
1  begin
2      foreach q_i ∈ Q do
3          while E_C(q_i) ≠ ∅ do
4              RSA(q_i, T_P, T_C, E_P, E_C, RSKY, SKY) → RSKY(q_i);
```

## VI. REVERSE SKYLINE PROCESSING FOR MULTIPLE QUERY POINTS

Solving the k-MAC problem requires processing multiple candidates, in order to first determine their influence sets. Subsequently, the algorithm presented in Section V can determine the solution of the k-MAC problem.

**A baseline solution.** A natural way to process multiple candidates would be to apply either the BRS or the RSA algorithm for each candidate individually, as shown in Alg. 4. We will refer to these approaches as *basic*. The *basic* algorithm is very inefficient in terms of I/O accesses, because it requires accessing each entry $e_p$ or $e_c$ entry multiple times, as each entry may appear in the corresponding priority queue of several candidates.

**Our Batch-RSA algorithm.** In order to eliminate the drawbacks of the basic algorithm, we now introduce our Batch-RSA algorithm, which exploits I/O commonalities and offers shared processing among candidates. Batch-RSA utilizes in its core the RSA algorithm that we presented in Section III.

A key requirement for the efficient execution of Batch-RSA is the grouping of candidates in such a way that the candidates have a high probability of sharing multiple active entries in their $E_P$ and $E_C$ sets. If this is somehow possible to achieve, then each of these entries, when expanded at an iteration of its RSA subcomponent, will be read from disk just once, for all these entries.

It is important to note that we cannot safely assume that the sets of active entries, along with all the necessary data structures that Batch-RSA utilizes for their efficient processing, fit in main memory. Based on the memory capabilities of the hardware which executes the Batch-RSA algorithm, and worst case estimates of the amount of customer and product entries in the sets of active entries, let us assume that $B$ candidates (where $B \ll |Q|$) fit in main memory and will be

**Algorithm 5:** Batch-RSA

**Input**: $Q$ a set of candidates, $T_P$ R-tree on products, $T_C$ R-tree on customers
**Variables**: $E_P(q_i)$ heap of active product entries for $q_i$, $E_C(q_i)$ set of active customer entries for $q_i$, $RSKY(q_i)$ reverse skylines for $q_i$, $SKY(q_i)$ midpoint skylines of $q_i$, $B_j$ batches with $|B_j| = B$

```
1  begin
2      partition Q into ⌈|Q|/B⌉ batches → B_j;
3      foreach B_j do
4          while (RSKY(q_i) for all q_i ∈ B_j have not been found) do
5              selectCandidate → q_i;
6              if E_C(q_i) ≠ ∅ then
7                  BRSA(q_i, B_j, T_P, T_C, E_P(q_i), E_C(q_i), RSKY(q_i), SKY(q_i));
```

processed as a batch. Using worst case estimates does not have a severe impact in the performance of Batch-RSA since, as we explain later in this section and demonstrate experimentally, it is desirable to utilize fairly modest values (i.e., 10 to 20) for $B$. Larger values may result in joint processing of dissimilar candidates, thus incurring unnecessary processing cost.

Initially, Batch-RSA partitions the candidate set into $\lceil |Q|/B \rceil$ batches based on a locality preserving hashing function, such as the Hilbert space filling curve. Therefore, each batch tends to include only candidates that are close to each other in the multidimensional space. Then, a modified version of the RSA algorithm, which handles batch processing of candidates, is executed on each batch (Alg. 5). The candidates in each batch are processed in a round robin fashion (Line 5).

Batch-RSA proceeds similar to RSA, except that whenever an entry $e_x$ is expanded, all local priority queues in which $e_x$ appears are appropriately updated. Recall that, for each candidate, children nodes are first checked with the candidate's currently found skyline set, and with other children nodes, as we discussed in Section III-B. Only those nodes that are not pruned are pushed to the priority queue of the respective candidate. Alg. 7 details the pseudocode for batch updating the priority queues of the batch.

## VII. EXPERIMENTAL EVALUATION

We now evaluate the performance of our proposed methods using both real and synthetically generated data. All algorithms examined in our experiments were implemented in C++ and executed on a 3GHz Intel Core 2 Duo CPU with 4GB RAM running Debian Linux. The code for the BRS algorithm was provided to us by the authors of [24].

### A. Used Data Sets

We used a publicly available generator [1] in order to construct synthetic data sets. We distinguish three classes of data sets, based on the distribution of the attributes' values. In uniform (UN) data sets, attribute values are drawn from a uniformly random distribution. In anti-correlated (AC) data sets, points having desirable attribute values in one dimension are more likely to have less desirable values in the other dimensions (e.g., laptops with faster CPU are more likely to be more expensive). Finally, in correlated (CO) data sets, points with desirable attribute values in one dimension tend to have good values in other dimensions as well.

## Algorithm 6: BRSA

**Input**: $B$ a set of candidates, $T_P$ R-tree on products, $T_C$ R-tree on customers, $E_P(q_i)$ heap of active product entries for $q_i$, $E_C(q_i)$ set of active customer entries for $q_i$, $RSKY(q_i)$ reverse skylines of $q_i$, $SKY(q_i)$ midpoint skylines of $q_i$
**Output**: $RSKY(q_i)$ reverse skylines of $q_i$

1 **begin**
2  **while** $E_C \neq \varnothing$ **do**
3   *dominated* := *false*;
4   $E_C(q).pop() \to e_c$;
5   **if** *dominated($e_c$, SKY($q_i$))* **then**
6    *dominated* := *true*;
7    *continue*;
8   **if** $e_c$ *is a non-leaf entry* **then**
9    BatchExpand($e_c, B, SKY(q_i), E_C(q_i), T_C$);
10   **else**
11    **foreach** $e_p \in E_P(q)$ **do**
12     *midpoint($e_p, q_i$)* $\to m$;
13     **if** $e_c$ *is dominated by* $m$ **then**
14      **if** $e_p$ *is a leaf entry* **then**
15       **if** *(dominated(m, SKY($q_i$))* == *false*) **then**
16        $SKY(q_i).push(m)$;
17       *dominated* := *true*;
18       *break*;
19      **else**
20       BatchExpand($e_p, B, SKY(q), E_p(q), T_P$);
21     $E_P(q_i).remove(e_p)$;
22   **if** *(dominated* == *false)* **then**
23    $RSKY(q_i).push(e_c)$;
24  **return** $RSKY(q_i)$;

## Algorithm 7: BatchExpand

**Input**: $e_x$ a product or customer entry, $B$ a set of candidates, $SKY(q_i)$ currently found midpoint skyline set w.r.t to $q_i$, $E_x(q_i)$ priority queue on $e_x$, $T_x$ index on $e_x$
**Output**: $E_x$
**Variables**: $L_i$ queues on children of $e_x$

1 **begin**
2  **foreach** $q_i \in B$ **do**
3   $L_i := \varnothing$;
4  **foreach** $e_{x_l}$ *children of* $e_x$ **do**
5   **if** $e_x$ *is a product entry* **then**
6    **foreach** $q_i \in B$ **do**
7     **if** *(dominated($e_{x_l}, L_i$)* == *true*) **then**
8      *continue*;
9   **foreach** $q_i \in B$ **do**
10    **if** *(dominated($e_{x_l}, SKY(q_i))$* == *false*) **then**
11     $L_i.push(e_{x_l})$;
12  **foreach** $q_i \in B$ **do**
13   **foreach** $e_{x_l} \in L_i$ **do**
14    $E_x(q_i).push(e_{x_l})$;
15  **return** $E_x$;

We also compared our algorithms on two real world data sets. The NBA data set (NBA) consists of 17,265 5-dimensional points, representing a player's annual performance. The dimensions are average values for the number of points scored, rebounds, assists, steals and blocks. The household data set (HOUSE), consists of 127,930 6-dimensional points, representing the percentage of an American family's annual income spent on 6 types of expenditure: gas, electricity, water, heating, insurance, and property tax. We used the real data sets as products (NBA-P and HOUSE-P resp.). Both data sets were normalized to have a range [0,1000] on each dimension. In order to construct customer (NBA-C, HOUSE-C) and candidate sets (NBA-Q, HOUSE-Q), for each point in NBA-P (HOUSE-P resp.) we added a gaussian noise with mean equal to the value of the actual point for the respective dimension and variance equal to 5 and 10 respectively. For each data set, we built an R-tree with a page size of 4K.

### B. Algorithms Used

We compared the performance of the BRS algorithm with our RSA algorithm and our Batch-RSA algorithm for computing the answer to k-MAC queries. Please recall that neither BRS nor RSA perform any batching of the examined candidates, while Batch-RSA presorts the new candidate products (candidates) and assigns them to partitions (batches).

In order to add another interesting baseline solution against RSA and Batch-RSA, we also devised and implemented a branch-and-bound algorithm, presented in Alg. 8, for the k-MAC problem. The BB-RSA maintains for each candidate, apart from its influence set, the currently found upper bound of its influence score. BB-RSA associates with each $q \in Q$ an R-tree $T_P(q)$ on the products and an aggregate (count) aR-tree [16] $T_C^A(q)$ on the customers. The aR-tree maintains $e_c.count$, i.e. the number of leaf entries that lie within the subtree of $e_c$, for each node $e_c$. To understand the use of the aR-tree, consider the case when a $e_c$ entry can be pruned. With a regular R-tree on the customers, we would have to access the pruned entry and enumerate the customer points that lie within $e_c$'s subtree in order to update the upper bound of the influence score. With an aR-tree this step can be avoided.

## Algorithm 8: BB-RSA

**Input**: $Q$ a set of candidates, $T_P$ R-tree on products, $T_C$ aR-tree on customers, $k$
**Output**: $Q_k$ min-heap on $IS^+$ containing the most attractive candidates, $|Q_k| = k$
**Variables**: $E_P(q_i)$ heap of active product entries for $q_i$, $E_C(q_i)$ heap of active customer entries for $q_i$, $RSKY(q_i)$ reverse skylines for $q_i$, $SKY(q_i)$ midpoint skylines of $q_i$, $IS^+(q_i)$ upper bound for $IS(q_i)$, $H_{max}$ max-heap on $IS^+$, $IS(Q_k)$ current lower bound of joint influence score

1 **begin**
2  $Q_k := \varnothing$;
3  **foreach** $q_i \in Q$ **do**
4   $RSA(q_i) := \varnothing; IS^+(q_i) := |C|; E_P(q_i) := \varnothing; E_C(q_i) := \varnothing$;
5  **while** $Q \neq \varnothing$ **do**
6   $H_{max}.pop() \to q_i$;
   /* Process $q_i$ if IS($q_i$) not completely refined */
7   **if** $E_C(q_i) \neq \varnothing$ **then**
8    BRSA($q_i, T_P, T_C, E_P, E_C, RSKY, SKY, IS^+, IS^-$);
9    **foreach** $q_i \in Q$ **do**
     /* Run kGSA using $IS^-$ for each candidate, to get $IS^-(Q_k)$ */
10    kGSA $\to Q_k^-$;
     /* Run kGSA using $IS^+$ for each candidate, to get $IS^+(Q_{k-1})$. Force kGSA to start by selecting $q_i$. */
11    kGSA $\to Q_{k-1}$;
12    **if** $IS^+(q_i) + IS^+(Q_{k-1}) < IS^-(Q_k)$ **then**
13     $Q := Q - \{q_i\}$; //Prune $q_i$
14  **return** $Q$;

BB-RSA proceeds in iterations and is executed similar to Batch-BRS with one main exception. If a customers' entry $e_c$ can be pruned, BB-RSA does not access $e_c$; instead the aggregate information $e_c.count$ is subtracted from $IS^+(q)$. Further, BB-RSA picks one candidate at a time (Line 6), performs a single execution of Batch-RSA (Line 8) and updates $IS^+(q)$.

In each iteration, BB-RSA selects to examine the candidate $q_i$ with the highest current $\text{IS}^+(q_i)$, and processes the candidate, unless the candidate is completely refined (Line 7). Finally, BB-RSA prunes $q_i$ based on the following pruning rule: Let $\text{IS}^-(Q_k)$ denote the lower bound of the currently optimal solution based on Alg. 3. Let $\text{IS}^+(Q_{k-1})$ denote the upper bound of the optimal solution based on Alg. 3, when forcing the algorithm to include $q_i$ in the solution. If the condition in Line 12 of BB-RSA is satisfied, then $q_i$ can safely be pruned, even if its exact influence set has not been determined.

The BB-RSA algorithm is introduced and compared against, since a branch and bound technique seems natural for the k-MAC problem. However, it has several drawbacks. First, BB-RSA cannot partition the candidates in batches, unless it aims to solve the 1-MAC problem. To understand why this is the case, let us consider the following scenario of seeking the solution to 2-MAC. Let us assume a candidate $q_i : |RSKY(q_i)| = 1$, and that the customer that $q_i$ has in its reverse skyline does not appear in the skyline of any other node. Then, it is not hard to devise cases where the solution to 2-MAC may involve combining $q_i$ with another candidate $q_j$ that has the largest influence score. Thus, if we want to enable BB-RSA to perform pruning, then all candidates need to be processed in a single batch. However, with this setup, the performance of BB-RSA was disappointing (this point will be made clear when we also examine the performance of Batch-RSA when using very large batch sizes).

We thus only present experiments for the 1-MAC case. This scenario is a best case scenario for the BB-RSA algorithm, since it allows its implementation using batching of the candidates (similarly to Batch-RSA) and also allows the pruning of candidates. It is important to emphasize that the BRS, the RSA and the Batch-RSA algorithms are not influenced by the value of $k$, since they first determine the influence sets of all candidates, and then run Alg. 3 (which requires a few milliseconds to run) to produce the final solution.

### C. Metrics and Default Parameter Setting

In each experimental setup, we vary a single parameter while setting the remaining to their default values. Table II summarizes the parameters under investigation and their corresponding ranges; the default values are depicted in bold.

TABLE II
PARAMETERS AND VALUES

| Parameter | Range |
|---|---|
| P dataset cardinality ($|P|$) | 10K, **100K**, 1M, 5M |
| C dataset cardinality ($|C|$) | 10K, **100K**, 1M, 5M |
| Q dataset cardinality ($|Q|$) | 100, **1K**, 10K |
| Dimensionality ($D$) | 2, **3**, 4, 5 |
| Batch Size ($|B|$) | 5, **10**, 20, 50, 100 |

Each set of experiments examines the performance of all methods in k-MAC query evaluation. In particular, we measure (*i*) the number of I/O operations (either on product or on customer entries), and (*ii*) the actual query processing time, consisting of the time spent both on CPU, as well as the time spent on I/O operations. Recall that single point skyline and batch algorithms require an additional step that greedily selects the best candidates after identifying the influence set of each candidate. However, the processing time required for this step is negligible (a few milliseconds) using Alg. 3 compared to the time required for the algorithm execution. Please note that the experimental results for the BRS and the RSA algorithms can also be used for evaluating their performance in single point reverse skyline query processing, since these algorithms simply process each candidate one after the other. Essentially, for both BRS and RSA algorithms, the plots represent the total CPU time and I/O operations required for processing a workload of $|Q|$ reverse skyline queries.

### D. Experimental Results

**Sensitivity Analysis vs.** $|P|$. We first performed a sensitivity analysis with respect to the size of the product data set. As depicted in Figures 5(a)-5(c), in terms of CPU cost, RSA is notably more efficient than BRS, particularly as the size of $|P|$ grows large. For instance, for uniformly distributed data with $|P| = 5M$ (Fig. 5(c)), BRS entails ∼8 times larger processing cost than RSA. With respect to the total I/Os required (Figures 5(d)-5(f)), RSA exhibits equivalent performance with BRS in the case where product and customer data have the same size (100K both). However, as the size of products increases, the I/Os required by RSA are significantly fewer than those of BRS. Notice the different behavior of the two algorithms with respect to the type of I/Os; BRS performs fewer accesses on the product index, whereas RSA is more efficient on customer I/Os. This is expected since the two algorithms follow different visiting orders. However, as demonstrated in this set of experiments, the strategy followed by RSA is more efficient in terms of the total number of I/Os. Further, both batch and branch-and-bound extensions of RSA achieve significant performance gains in all settings. Essentially they remain largely unaffected by the size of $|P|$. BB-RSA is slightly more expensive than Batch-RSA, since it requires more index space for the same size of input in order to maintain the aggregate information, and hence it entails more I/Os than Batch-RSA on the customer index.

**Sensitivity Analysis vs.** $|C|$. Next, we compared the efficiency of our algorithms with respect to the size of the customer data set. Figures 6(a)-6(c) depict the CPU cost, while Figures 6(d)-6(f) plot the I/O cost. As illustrated, RSA is slightly more efficient in terms of the CPU spent, whereas RSA and BRS require roughly the same number of I/Os. To justify this behavior, recall that RSA processes customers in iterations, hence the number of iterations grows with $|C|$. In contrast, in each iteration of BRS multiple $e_c$ entries might be processed. However, even in this (worst case for RSA) experimental setup, RSA performs equivalently to the BRS algorithm. Finally, for k-MAC queries both Batch-RSA and BB-RSA algorithms are significantly more efficient from the single point algorithms.

**Sensitivity Analysis vs.** $|Q|$. We now investigate the performance of all algorithms in evaluating k-MAC queries with respect to the size of the candidates data set. Figures 7(a)-7(f) plot the results in logarithmic scale. Recall that for both BRS and RSA plots represent the total time (I/Os resp.) required for a workload of $|Q|$ reverse skyline queries, which are processed one after the other, whereas Batch-RSA and BB-RSA proceed in batches. An interesting observation comes when analyzing the CPU cost of Batch-RSA and BB-RSA in comparison to single point algorithms. Batch-RSA and BB-RSA both require more CPU cost than RSA, because of the time spent on maintaining priority queues for all candidates in the batch. However, their total processing cost is less because of the reduced number of required I/Os. Overall, Batch-RSA is the most efficient and scalable algorithm for the k-MAC query when the size of $|Q|$ increases.

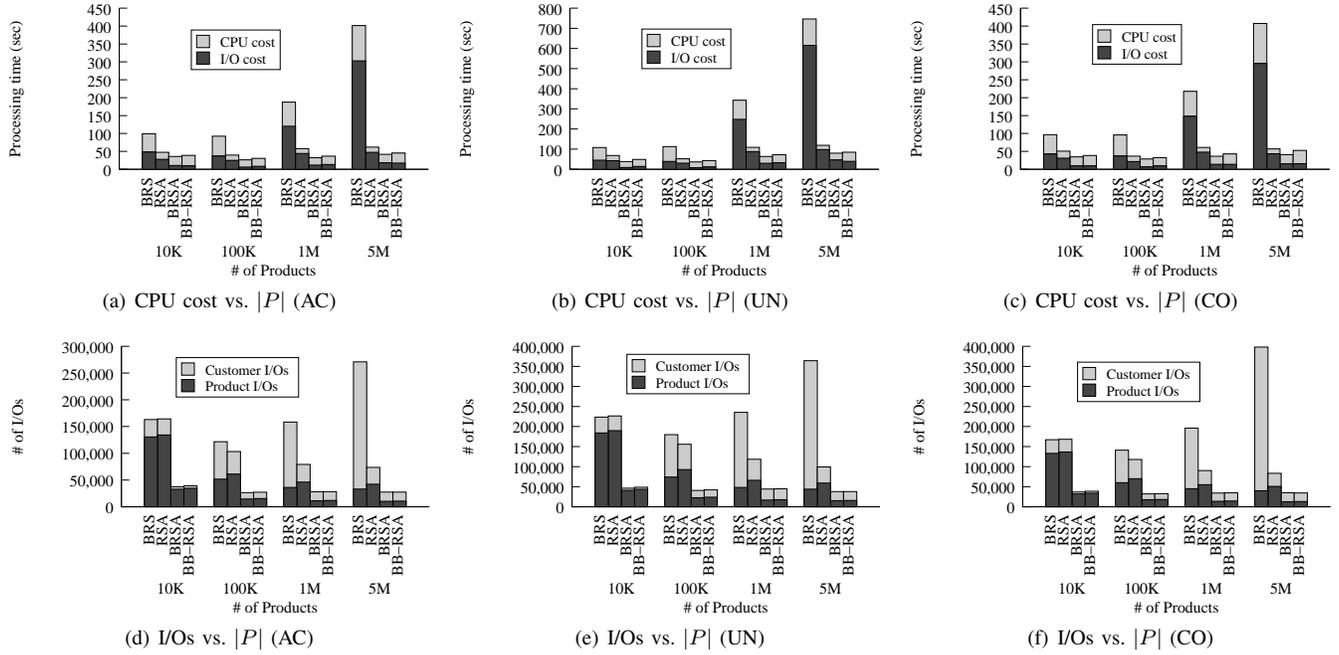

Fig. 5. Varying number of products $|P|$

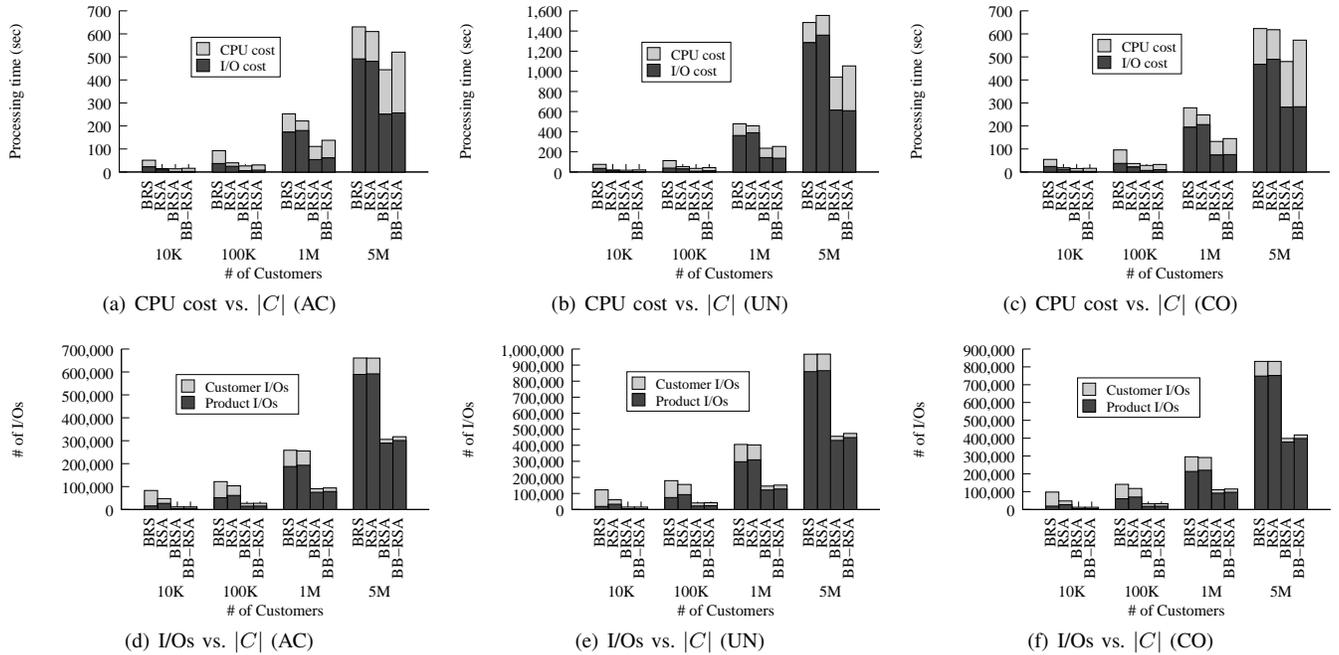

Fig. 6. Varying number of customers $|C|$

**Sensitivity Analysis vs. $D$.** Next, we varied the dimensionality of the data sets from 2 to 5 and investigated the performance of all algorithms with respect to $D$. Figures 8(a)-8(c) and 8(d)-8(f) plot the results for the CPU and I/O costs respectively. As expected (refer to the complexity analysis in Section III-A), in higher dimensional data BRS requires one order of magnitude more processing time to complete than RSA. Clearly, as shown in Figures 8(a)-8(c), in higher dimensions the I/O cost of BRS is dominated by the processing cost (notice that plots are in logarithmic scale). To understand this, recall that each dominance check requires $D$ comparisons, and that the number of minmax points per entry is $D$.

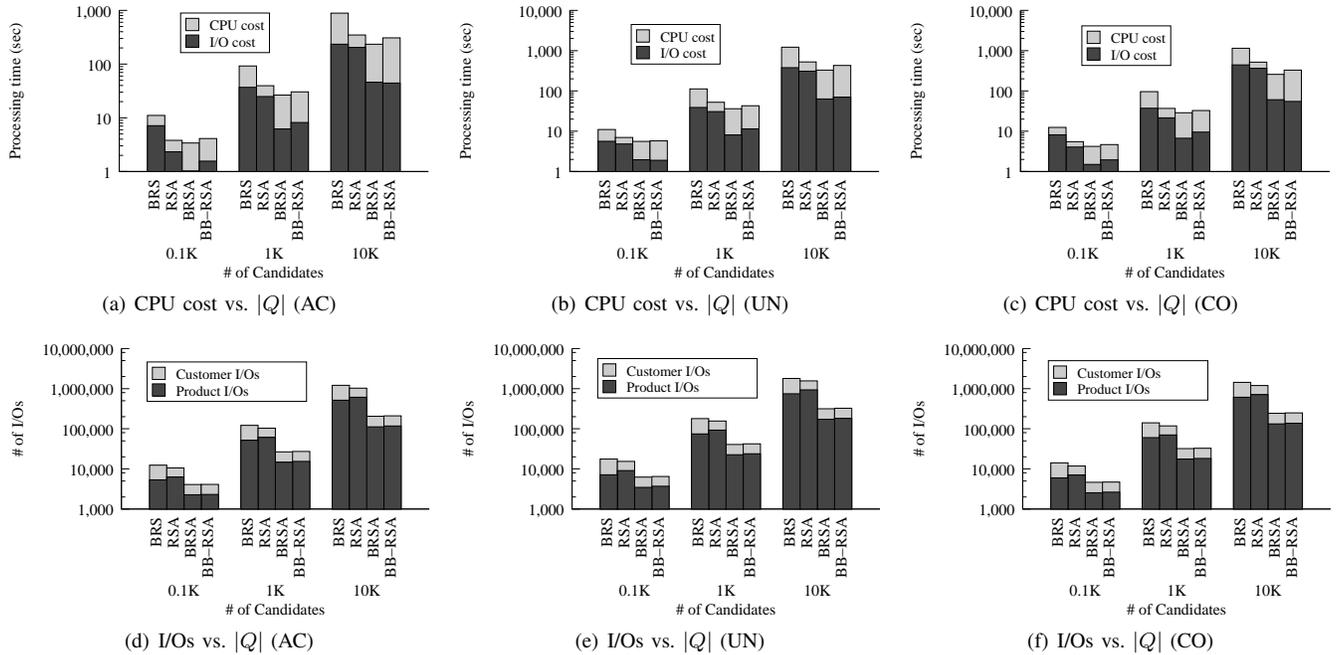

Fig. 7. Varying number of candidates $|Q|$

**Sensitivity Analysis vs.** $|B|$. We now examine the batch and branch-and-bound algorithms in respect to the batch size $|B|$. Recall that each candidate is hashed to the appropriate batch, such that each batch contains only candidates that are close in the initial space. Therefore, we expect that candidates in the same batch will share pruning of product or customer entries. We set each batch to contain from 5 to 100 candidates and plotted the results in Figures 9(a)-9(f). As depicted, as batch size increases, the total I/O operations fall since more pruning can be shared among candidates. However, the total processing cost grows rapidly. This happens due to the following reason. When the batch size is larger, candidates belonging to the batch are dissimilar, and hence they will not benefit from the expansion of entries from other candidates. Rather, children entries have to be pushed into the local priority queues of most of the candidates. This results to: (*i*) increasing cost and larger space required for the maintenance of $E_P(q_i)$ and $E_C(q_i)$, and (*ii*) more processing time for subsequent dominance checks. As showcased by the experiments, keeping smaller batch sizes (up to 10 candidates) positively favors the performance of both Batch-RSA and BB-RSA. At this point it is important to emphasize that keeping all candidates in a single batch is impractical; this fact restrains BB-RSA algorithm for being used for k-MAC queries other than 1-MAC.

**Progressiveness**. Finally, we compared the progressiveness of RSA and BRS algorithms. This is important for applications that require a quick response or when the complete output is not essential. We executed both algorithms on a workload of $|Q|$ queries using the same default settings (refer to Table. II). Specifically, the x-axis represents the percentage of reverse skyline results found so far with regards to the total influence score. The y-axis plots the time required to report the corresponding percentage of results, both in absolute values (Figures 10(a), 10(e)) (Figures 10(b), 10(f)).The figures clearly demonstrate the superiority of RSA algorithm in terms of progressively reporting the query results. In particular, RSA requires one order of magnitude less time to output the first 5% of results.

**Experiments with Real Data**. Figures 11(a)-12(d) report our experimental findings on the real data sets used. The results are in accordance with our experimental study on synthetic data sets. However, notice that the differences in performance between RSA and BRS algorithms are more intense in real data sets (especially vs. $|C|$) due to the high dimensionality. Further, there was no point to include a sensitivity analysis vs. the number of candidate products $|Q|$ for HOUSE data set, since BRS algorithm required 43 hours to run on our system for $|Q|$ = 1K. These facts further justify our motivation for a more efficient and scalable algorithm for reverse skyline queries. Overall, RSA outperforms BRS for the single point reverse skyline query, whereas Batch-RSA is the best choice for k-MAC queries, in all experimental settings tested.

## VIII. RELATED WORK

The skyline query returns the set of not dominated objects, corresponding to the Pareto optimal set. [3] introduce skyline queries in the context of database research, also presenting various external memory algorithms. Since then, several non-indexed [5], [9], [2] and indexed-based algorithms [12], [17], [13] have been presented in skyline research literature.

Assuming a set of products $P$, the dynamic skyline [17] contains all products $p \in P$ that are 'attractive' according to a customer's preferences. In order to capture the manufacturer's perspective, [6] introduces reverse skyline query, which returns all customers $c \in C$ that would find a product $p$ as 'attractive', and proposes a branch and bound extension of the BBS algorithm [17] that reduces the search space. [15] improves upon [6] by providing tighter pruning rules based on midpoint skylines (Section II-B) and proposes algorithms for calculating reverse skylines on uncertain data. [24] introduces the BRS algorithm (Section II-C) which exploits some additional optimizations that are possible for the case of certain data.

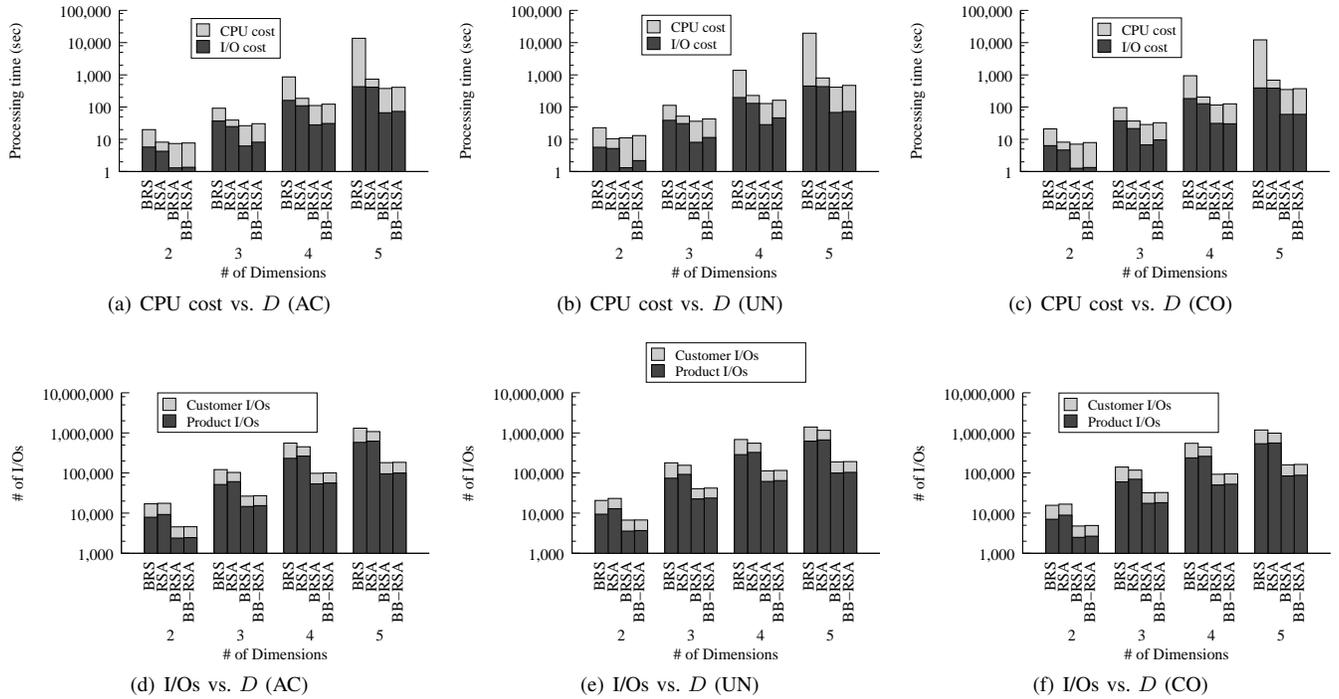

Fig. 8. Varying number of dimensions $D$

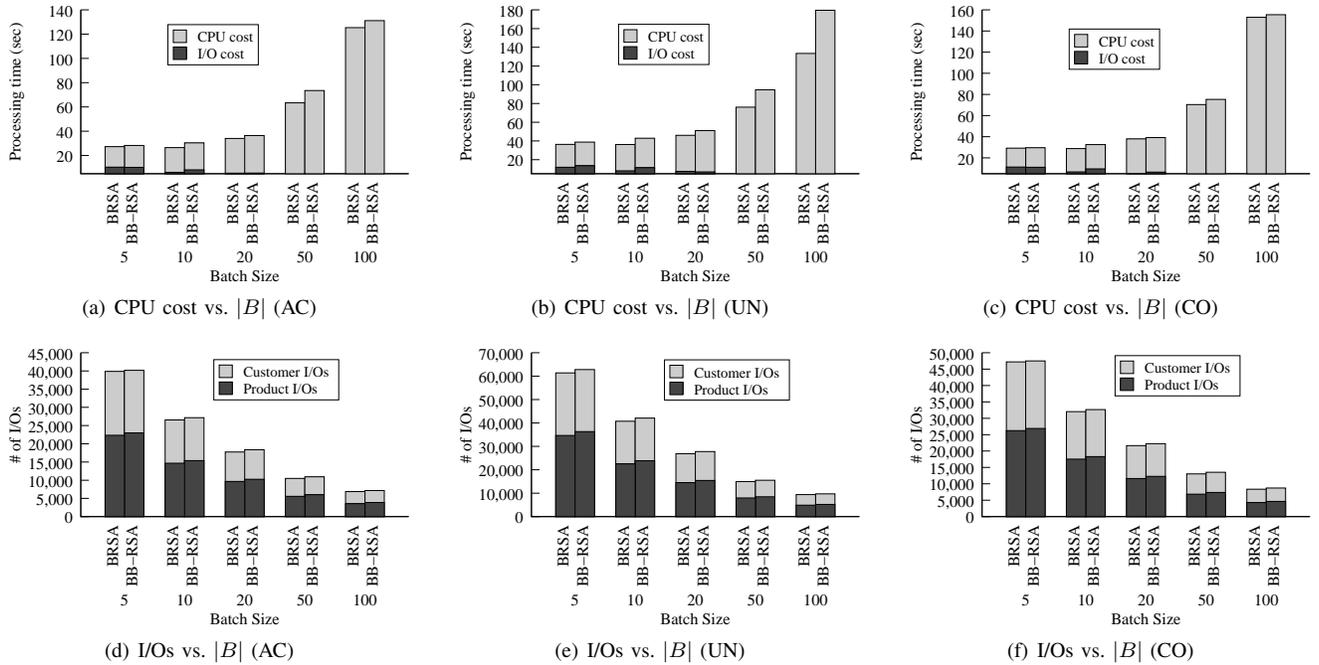

Fig. 9. Varying batch size $|B|$

Several other skyline-based approaches that focus on the manufacturer's view have been proposed as well. DADA [14] aims to help manufacturers position their products in the market, based on dominance relationship analysis. A product $p$ satisfies a customer $c$ iff $p$ dominates $c$. However, DADA does consider user preferences, hence it can only be used on static dimensions, i.e. those having optimal values that are independent of user preferences. Another approach similar in motivation to our work is [21] that addresses the problem of identifying competitive packages formulated by combining individual products, such as a flight and a hotel. A package is competitive if it is not dominated by any other package. In [22] the authors extend this work by proposing methods to calculate the $k$ most profitable products, i.e. those that maximize the profit gain measured as the package price. Apart from the different application setting from our work, [21], [22] do not consider dynamic dimensions or multiple customers either. In contrast, our work captures the profitability of each product in terms of the number of potential buyers.

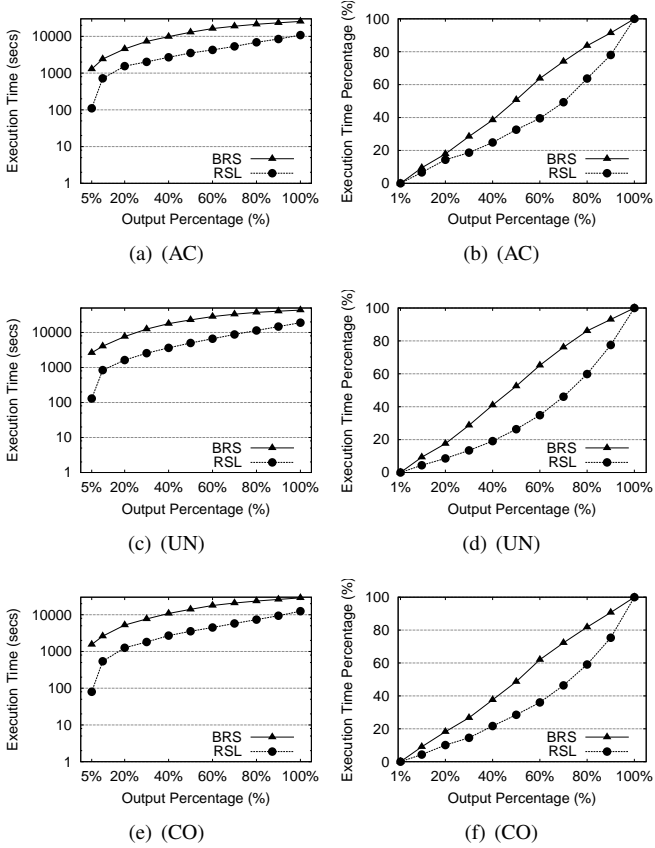

Fig. 10. Progressiveness

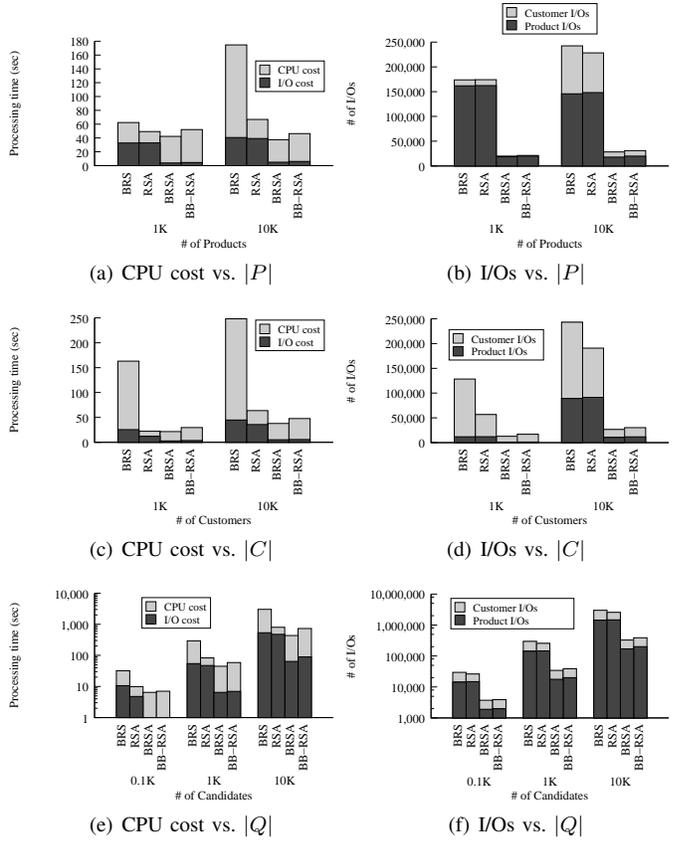

Fig. 11. Experiments - NBA Dataset

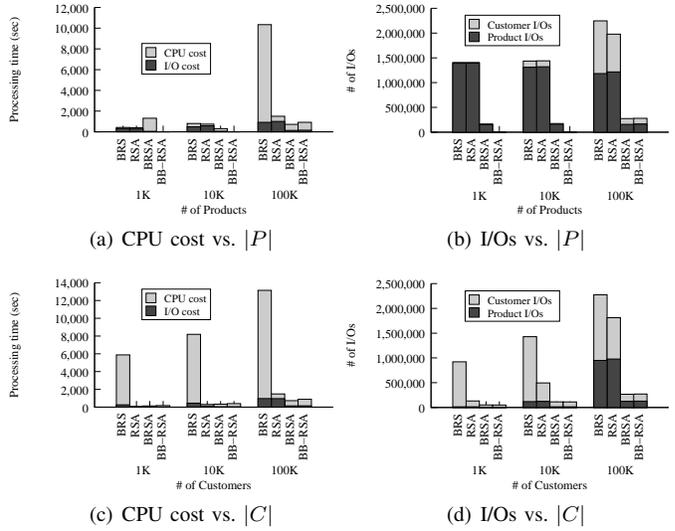

Fig. 12. Experiments - HOUSEHOLD Dataset

Assuming that user preferences are formulated as functions, the skyline set will always include the top-1 result for any preference function that is monotone. Due to the difficulty to come up with a precise preference function in many scenarios, skyline queries have been particularly useful in preference queries. Due to the difficulty to come up with a precise preference function in many scenarios, skyline queries have been particularly useful in preference queries. On the other hand, assuming that a linear preference function is available, *top-k queries* [4], [7] provide an alternative solution to skylines.

Given a product $p$ and set of user preference functions $W$ formulated as weight vectors, [19] introduces *reverse top-k queries*. In analogy to reverse skylines, the reverse top-k query returns those weight vectors such that $p$ belongs to their top-k query results. [20] extends [19] by addressing the problem of discovering the top-m products that maximize the cardinality of their reverse top-k set. Although the spirit of [19] is similar to our approach, it requires learning and precisely formulating user preference functions as weight vectors, which is rather impractical in real life. On the other hand, in our work each user is allowed to formulate her preferences directly in terms of the desired product specifications. Moreover, in contrast to our k-MAC query, the total profitability of the query result in [19] is not guaranteed because multiple products might be attractive for the same potential buyers.

Finally, the skyline query exhibits some similarities with the *nearest neighbor query* (NN) [18]. spatial databases. Given a set of spatial objects (e.g. facilities) $P$ and a query point $q$ (e.g. a user's spatial location), the $k$-*nearest neighbor query* ($kNN(q)$) contains the $k$ objects $p \in P$ having the minimum spatial distance to $q$. Further, given an object $p$ and a set of customers $C$, the *bichromatic reverse nearest neighbor*

*query* (RNN($p$)) [11] returns the customers $c \in C$ whose NN contains $p$. The work in [25] selects the $k$ objects from a set $P$ inside a given spatial region $Q$, such that they have the largest RNN set. Finally, a more recent work [23] addresses the problem of identifying the region to establish a new facility so that the size of its RNN set is maximized. *vantage points*, so that the most influential among them define the optimal region. The problem formulation and the proposed solutions are similar to our approach. However none of these methods can be directly applied to our setting, because the 'attractiveness' of each product $p$ is based on its dominance relationships with other competitor products with respect to a customer $c$, rather than its distance from $c$.

**Acknowledgements**. This research has been co-financed by the European Union (European Social Fund - ESF) and Greek national funds through the Operational Program "Education and Lifelong Learning" of the National Strategic Reference Framework (NSRF) - Research Funding Program: Heracleitus II. Investing in knowledge society through the European Social Fund.

## IX. CONCLUSIONS

In this work, we studied two important class of queries involving customer preferences. We first proposed a novel algorithm, RSA, for reverse skyline query evaluation. We then introduced a new type of skyline query, termed as the $k$-Most Attractive Candidates (k-MAC) query; k-MAC selects the set of $k$ candidate products that jointly maximizes the total number of expected buyers. We propose a batch algorithm, which utilizes in its core our RSA algorithm, but significantly improves upon naive techniques that process all candidate products individually, for answering k-MAC queries. Our extensive experimental study on both real and synthetic data sets demonstrates that (*i*) RSA outperforms the BRS algorithm, for the reverse skyline problem, in terms of I/O, CPU cost and progressiveness of output, and (*ii*) that our proposed batch algorithm outperforms branch-and-bound and baseline approaches for the k-MAC query.